\documentclass[runningheads]{llncs}
\usepackage{amssymb}
\usepackage{amsmath}
\usepackage{graphicx}
\usepackage{url}
\usepackage{nicefrac}
\usepackage{url}
\usepackage{bm}
\usepackage[normalem]{ulem}
\usepackage{tabularx}
\usepackage{subcaption}
\usepackage{adjustbox}

\usepackage{array}
\usepackage{booktabs}

\usepackage{harmony}

\usepackage{tikz}
\usepackage{pgfplots}
\pgfplotsset{compat=1.3}
\usepackage{filecontents}
\usetikzlibrary{matrix}
\usetikzlibrary{calc}
\usetikzlibrary{arrows}
\usetikzlibrary{patterns}
\usetikzlibrary{plotmarks}
\usetikzlibrary{intersections} 
\usetikzlibrary{decorations.pathmorphing}
\usetikzlibrary{decorations.pathreplacing}
\usetikzlibrary{decorations.markings,arrows} 
\usetikzlibrary{decorations.shapes}
\usetikzlibrary{fit}
\usetikzlibrary{backgrounds}
\usetikzlibrary{trees}

\newcommand{\set}[1]{\left\{ #1 \right\}}

\newcommand{\lossce}[1]{\mathcal{L}_{\mathrm{CE}}\!\left(#1\right)}

\usepackage[colorinlistoftodos]{todonotes}
\usepackage{soul}

\begin{document}
\title{Rhythm, Chord and Melody Generation for Lead Sheets using Recurrent Neural Networks}
\titlerunning{Rhythm, Chord and Melody Generation for Lead Sheets using RNNs}
%
\author{Cedric De Boom\and Stephanie Van Laere\and Tim Verbelen\and Bart Dhoedt}
\authorrunning{C.~De Boom, et al.}
%
\institute{IDLab, Department of Information Technology at Ghent University -- imec\\Technologiepark-Zwijnaarde 126, B-9052 Ghent, Belgium\\ \email{firstname.lastname@ugent.be}}
\maketitle              
\begin{abstract}
Music that is generated by recurrent neural networks often lacks a sense of direction and coherence.
We therefore propose a two-stage LSTM-based model for lead sheet generation, in which the harmonic and rhythmic templates of the song are produced first, after which, in a second stage, a sequence of melody notes is generated conditioned on these templates.
A subjective listening test shows that our approach outperforms the baselines and increases perceived musical coherence.

\keywords{Music generation \and lead sheets \and neural networks}
\end{abstract}

\section{Lead Sheets}
\label{sec:leadsheets}
Lead sheets are widely used to represent the fundamental musical information about almost any contemporary song: they contain a chord scheme, a melody line, some navigation and repetition markers, and sometimes lyrics.
They seldom contain information about the instrumentation or accompaniment, so any band can take a lead sheet as a guideline and make the song their own, sometimes even by improvising over the chord schemes.
In this paper we focus on generating chords and melody lines for lead sheets from scratch.

A major difficulty in music generation is that harmony, melody and rhythm all influence each other.
For example, a melody note can change whenever the underlying harmony changes, and vice versa.
Rhythmic patterns can influence which notes are played, and rhythm and harmony together define the overall groove of the piece.
To tackle this issue, we split the generation process into two stages.
First, we generate a harmonic progression using chord sequences, while simultaneously picking the most appropriate rhythmic patterns.
And in a second step the melody is generated on top of this harmonic and rhythmic template.

We are, however, not the first to tackle the problem of lead sheet generation and, in general, music generation.
Briot et al.~provide a recent and extensive overview of all deep learning based techniques in this field \cite{Briot:2017ts}.
Regarding lead sheets specifically, Liu et al.~use GAN-based models on piano roll representations, but the melody and chords are still predicted independently by different generators \cite{Liu:2018tz}.
Roy et al.~devise a lead sheet generator with user constraints defined by Markov models, and harmonic synchronization between melody and chords through a probabilistic model that encodes which melody notes fit on which chords \cite{Roy:2017ui}.
There have also been many efforts in the past that learn to generate chords for a given melody \cite{Lim:2017td,Pachet:2014vg,Huang:2018td} or the other way round \cite{Yang:2017ve}.
In this paper we want to show that, on the one hand, a two-stage generation process greatly improves the perceived quality of the music.
And, on the other hand, we show that melodic coherence improves when the melody generator gets to look ahead at the entire harmonic template of the song.

\begin{figure*}[t]
\centering
\includegraphics[trim = 1cm 26.5cm 14.5cm 1cm, clip, width=0.35\textwidth]{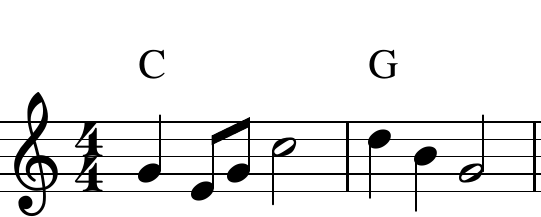}
\qquad\qquad
\tabcolsep=0.2cm
\scriptsize
\begin{tabular}{c | c c c}
\toprule
$i$ & $c_i$ & $r_i$ & $m_i$ \\
\hline
1 & C & quarter & G4 \\
2 & C & eighth & E4 \\
3 & C & eighth & G4 \\
4 & C & half & C5\\
5 & $\mid$ & $\mid$ & $\mid$ \\
6 & G & quarter & D5 \\
7 & G & quarter & B4 \\
8 & G & half & G4 \\
9 & $\mid$ & $\mid$ & $\mid$ \\
\bottomrule
\end{tabular}
\caption{Example of a lead sheet decomposition into chords, rhythms and melodies.}
\label{fig:example_lead_sheet}
\end{figure*}

We formally define a lead sheet $x_{1:n}$ of length $n$, characterized by a sequence of chords $c_{1:n}$, rhythms $r_{1:n}$ and melody pitches $m_{1:n}$.
At each time step $i$ in the piece, each of these quantities take some value:
\begin{align}
    x_{1:n} &= \set{ c_{1:n}, r_{1:n}, m_{1:n} },\quad x_i = \set{ c_i, r_i, m_i }.
\end{align}
In this equation, the compact $x_{j:k}$ notation denotes the sequence $\left( x_j, x_{j+1} \dots x_k\right)$.
Figure \ref{fig:example_lead_sheet} shows an example of this decomposition.
Notice that the chords are repeated until there is a change in harmony, thereby allowing us to model the entire lead sheet using a single shared time scale.
We also point out that there is only one melody note per time step, which is not a severe restriction, since most lead sheets only contain monophonic melodies.
Finally, we choose to treat the barlines as separate elements in the sequence, which is indicated by the vertical bars in Figure \ref{fig:example_lead_sheet}.

\section{The Wikifonia Dataset}
In this paper we will make use of the Wikifonia dataset, a former public lead sheet repository hosted by wikifonia.org.
It contains more than 6,500 lead sheets in MusicXML format, and in all sorts of modern genres.
This section goes over the different preprocessing and encoding steps that are executed on the dataset in order to obtain a clean collection of lead sheets.

\subsection{Preprocessing}

\paragraph{Eliminate polyphony} 
Whenever multiple notes sound at the same time, we only retain the note with the highest pitch, as it is often the note that characterizes the melody.

\vspace*{-1mm}
\paragraph{Ignore ties}
Connections between two notes with the same pitch that extend the first note's duration are ignored.
The two notes are therefore treated as two separate notes with their original duration.

\vspace*{-1mm}
\paragraph{Delete anacruses} Incomplete bars that often appear at the start of a piece, are removed from all lead sheets.

\vspace*{-1mm}
\paragraph{Unfold repetitions} 
Lead sheets can contain repetition and other navigation markers.
If a section should be repeated, we duplicate that particular section, thereby unfolding the piece into a single linear sequence.

\vspace*{-1mm}
\paragraph{Remove ornaments} 
Since such ornaments do not contribute much to the overall melody, we leave them out.

\vspace*{-1mm}
\subsection{Data encoding and features}
After preprocessing, we encode the melody, rhythm and chord symbols into feature vectors such that they can be used as input to our generators.

\vspace*{-1mm}
\paragraph{Encoding rhythms}
We retain the 12 most common rhythm types in the dataset, which are given in Appendix \ref{appendix:rhythmtypes}.
We remove 184 lead sheets from the dataset that contain other than these 12 types.
Together with the representation for a barline, we encode rhythm into a 13-dimensional one-hot vector $\vec{r}_i$.

\vspace*{-1mm}
\paragraph{Encoding chords}
A chord is described by both its root and its mode.
There are 12 possible roots (C, C$\sharp$, D, D$\sharp$, \dots, B) and we choose to convert all accidentals to either no alteration or one sharp.
We count 47 different modes in the dataset, which we map to one of the following four: major, minor, diminished or augmented.
This mapping only very slightly reduces musical expressivity and interestingness.
The mapping table can be found in Appendix \ref{appendix:modemapping}.
The 12 roots and 4 modes give 48 chord options in total, resulting in a 49-dimensional one-hot vector $\vec{c}_i$ if we include the barline.

\vspace*{-1mm}
\paragraph{Encoding melody}
The MIDI standard defines 128 possible pitches.
We assign two additional dimensions for rests and barlines, resulting in a 130-dimensional one-hot encoded melody vector $\vec{m}_i$.

\vspace*{-2mm}
\section{Recurrent Neural Network Design}
\begin{figure}[t]
\centering
\begin{adjustbox}{angle=90}
\begin{tikzpicture}[->,thin,>=stealth',yscale=-1]

    \node (input1) at (0,0) [draw,thin,minimum width=1.0cm,minimum height=0.2cm,text width=1cm,align=center] {\scriptsize $\vec{c}_i$};
    \node (input2) at (1.0,0) [draw,thin,minimum width=0.5cm,minimum height=0.2cm,text width=0.5cm,align=center] {\scriptsize $\vec{r}_i$};
    
    \node (lstm1) at (0.4,0.7cm) [draw,thin,minimum width=1.5cm,minimum height=0.3cm,text width=1.5cm,align=center] {\scriptsize LSTM 512};
    
    \node (lstm2) at (0.4,1.4cm) [draw,thin,minimum width=1.5cm,minimum height=0.3cm,text width=1.5cm,align=center] {\scriptsize LSTM 512};
    
    \draw[->,>=stealth'] (input1) to (lstm1);
    \draw[->,>=stealth'] (input2) to (lstm1);
    \draw[->,>=stealth'] (lstm1) to (lstm2);
    
    \draw[->,>=stealth'] (lstm1) to [out=197,in=175,looseness=3] (lstm1);
    \draw[->,>=stealth'] (lstm2) to [out=197,in=175,looseness=3] (lstm2);
    
    \node (dense) at (0.4,2.1cm) [draw,thin,minimum width=1.5cm,minimum height=0.3cm,text width=1.5cm,align=center] {\scriptsize Dense 61};
    
    \draw[->,>=stealth'] (lstm2) to (dense);
    
    \node (output1) at (0,2.8cm) [draw,thin,minimum width=1.0cm,minimum height=0.2cm,text width=1cm,align=center] {\scriptsize $\hat{\vec{c}}_{i+1}$};
    \node (output2) at (1.0,2.8cm) [draw,thin,minimum width=0.5cm,minimum height=0.2cm,text width=0.5cm,align=center] {\scriptsize $\hat{\vec{r}}_{i+1}$};
    
    \draw[->,>=stealth'] (dense) to (output1);
    \draw[->,>=stealth'] (dense) to (output2);

    \node (xinput1) at (0,4.2cm) [draw,thin,minimum width=1.0cm,minimum height=0.2cm,text width=1cm,align=center] {\scriptsize $\hat{\vec{c}}_{i+1}$};
    \node (xinput2) at (1.0,4.2cm) [draw,thin,minimum width=0.5cm,minimum height=0.2cm,text width=0.5cm,align=center] {\scriptsize $\hat{\vec{r}}_{i+1}$};
    
    \draw[->,>=stealth',dotted] (output1) to (xinput1);
    \draw[->,>=stealth',dotted] (output2) to (xinput2);
    
    \node (xbilstm1) at (0.375,4.9cm) [draw,thin,minimum width=1.75cm,minimum height=0.4cm,text width=1.75cm,align=center] {\scriptsize BiLSTM 1024};
    
    \node (xbilstm2) at (0.375,5.6cm) [draw,thin,minimum width=1.75cm,minimum height=0.4cm,text width=1.75cm,align=center] {\scriptsize BiLSTM 1024};
    
    \node (xinput3) at (1.78,5.6cm) [draw,thin,minimum width=0.5cm,minimum height=0.4cm,text width=0.5cm,align=center] {\scriptsize $\vec{m}_i$};
    
    \node (xlstm1) at (0.95,6.3cm) [draw,thin,minimum width=1.5cm,minimum height=0.3cm,text width=1.5cm,align=center] {\scriptsize LSTM 512};
    
    \node (xlstm2) at (0.95,7.0cm) [draw,thin,minimum width=1.5cm,minimum height=0.3cm,text width=1.5cm,align=center] {\scriptsize LSTM 512};
    
    \draw[->,>=stealth'] (xinput1) to (xbilstm1);
    \draw[->,>=stealth'] (xinput2) to (xbilstm1);
    \draw[->,>=stealth'] (xbilstm1) to (xbilstm2);
    \draw[->,>=stealth'] (xinput3) to (xlstm1);
    \draw[->,>=stealth'] (xbilstm2) to (xlstm1);
    \draw[->,>=stealth'] (xlstm1) to (xlstm2);
    
    \draw[<->,>=stealth'] (xbilstm1) to [out=196,in=174,looseness=3] (xbilstm1);
    \draw[<->,>=stealth'] (xbilstm2) to [out=196,in=174,looseness=3] (xbilstm2);
    \draw[->,>=stealth'] (xlstm1) to [out=197,in=175,looseness=3] (xlstm1);
    \draw[->,>=stealth'] (xlstm2) to [out=197,in=175,looseness=3] (xlstm2);
    
    \node (xdense) at (0.95,7.7cm) [draw,thin,minimum width=1.5cm,minimum height=0.3cm,text width=1.5cm,align=center] {\scriptsize Dense 130};
    
    \draw[->,>=stealth'] (xlstm2) to (xdense);
    
    \node (xoutput) at (0.95,8.4cm) [draw,thin,minimum width=0.5cm,minimum height=0.2cm,text width=0.7cm,align=center] {\scriptsize $\hat{\vec{m}}_{i+1}$};
    
    \draw[->,>=stealth'] (xdense) to (xoutput);
    
\end{tikzpicture}
\end{adjustbox}

\caption{The RNN architecture for both stage one (left) and stage two (right). The output dimensionality for every layer is written in each of the blocks. Whenever two blocks appear next to each other, the (output) vectors are concatenated.}
\label{fig:rnn_architectures}
\end{figure}
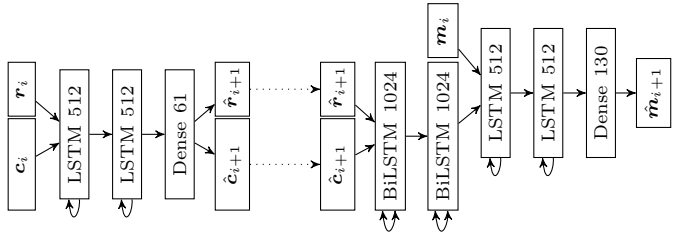

As mentioned in Section \ref{sec:leadsheets}, the lead sheet generation process happens in two stages: in stage one the rhythm and chord template of the song is learned, and in stage two the melody notes are learned on top of that template.
We will use separate LSTM-based models for both stages \cite{Hochreiter:1997fq}; the models are trained independently of each other, but they are combined at inference time to generate an entire lead sheet from scratch.
Figure \ref{fig:rnn_architectures} shows the complete architecture.

\paragraph{Stage one}
In this stage, the rhythm and chord vectors are first concatenated and are subsequently given as inputs to two LSTM layers followed by a dense layer.
All LSTM layers have a output dimensionality of 512 states, as indicated in the figure.
The output of the dense layer is cut in two vectors, both on which we apply a softmax nonlinearity with temperature $\tau$, controlling the concentration of the output distribution.
This way we are effectively modeling a distribution over the chord and rhythm symbols that come next in the sequence.

\paragraph{Stage two}
The second model will process the generated sequence of predicted chords and rhythms.
To this end, each chord and rhythm vector is again concatenated before being processed by two BiLSTM layers.
These BiLSTM states allow the pitch generator to look back and also ahead at the harmonic sequence, as inspired by \cite{Lim:2017td}.
The dimensionality of the BiLSTM layers is 512 in both directions, adding up to a total of 1024 states.
After concatenation with the previous melody vector, the BiLSTM states are fed through another stack of two LSTM layers.
The output of the last dense layer is used to predict the next melody note, again using a softmax nonlinearity controlled by a temperature parameter.

\subsection{Optimization details}
In this paper we train both stages separately; 
it is possible to jointly train both models through reparameterization tricks \cite{Jang:2016ub}, but we leave this as future work.
While training is done separately, inference of lead sheets from scratch is easily done by feeding the output of the first model to the input of the second model.
We use a standard cross-entropy loss function on all outputs.
In stage one we sum the losses on both the chord and rhythm outputs with hyperparameter $\alpha$:
\begin{align}
    \mathcal{L}_{\mathrm{stage\,1}}\!\left(\hat{\vec{c}}, \hat{\vec{r}}\right) = \alpha\cdot\lossce{\hat{\vec{c}}} + (1 - \alpha)\cdot\lossce{\hat{\vec{r}}}.
\end{align}
In this equation, $\lossce{\cdot}$ indicates the cross-entropy loss.
In stage two the loss function is equal to the cross-entropy loss on the melody output.
\begin{align}
    \mathcal{L}_{\mathrm{stage\,2}}\!\left(\hat{\vec{m}}\right) = \lossce{\hat{\vec{m}}}.
\end{align}
We use Adam with learning rate $\lambda$ to optimize both models \cite{Kingma:2015ku}.
During optimization, the training data is augmented by shifting the pitches and chords with a random number of semitones between -12 and 12, that is, between plus or minus one octave.
This virtually increases the amount of training data by a factor of 25, thereby aiding model generalization towards less frequently appearing keys.

\section{Experiments}

\paragraph{Hyperparameters}
In all experiments we use a batch size of 128 sequences, each of length 100.
The learning rate $\lambda$ is set fixed to 0.001.
We empirically found that a value $\alpha$ of 0.5 leads to good results, so we set it fixed to that value.
We also set the temperature $\tau$ slightly lower than 1 during inference of the melody, which helps to improve the perceived quality of the generated music; we varied it between 0.75 and 1.0 during the experiments.
For the rhythm and chord patterns a temperature of 1.0 gives the most pleasing results.

\vspace*{-1mm}
\paragraph{Baselines}
We will compare our model against two baselines:
\vspace*{-1mm}
\begin{enumerate}
    \item An unconditioned LSTM-based model similar to the stage one model in Figure \ref{fig:rnn_architectures}, but now the melody is also concatenated to the input and output.
    The melody is no longer conditioned on the entire chord and rhythm sequence.
    We also add an extra LSTM layer, adding up to a total of three.
    \item A two-stage model where the BiLSTM layers are replaced by regular LSTM layers, so that the melody cannot look ahead at the harmonic sequence.
    We keep all other parameters identical to the original model.
\end{enumerate}

\vspace*{-1mm}
\paragraph{Subjective listening test}
We conducted an online listening test in which we asked 40 participants to score 12 short audio clips, each of approximately one minute long.
The following songs were included in the test\footnote{Listen to the audio clips at \url{https://users.ugent.be/~cdboom/music}}:
\vspace*{-1mm}
\begin{itemize}
    \item 3 pieces generated by the two-stage model from scratch,
    \item 3 pieces generated by the two-stage model, but conditioned on the chord and rhythm scheme of existing songs: \textit{I Have a Dream} (Abba), \textit{Autumn Leaves} (jazz standard) and \textit{Colors of the Wind} (Alan Menken),
    \item 2 pieces generated by the one-stage baseline model,
    \item 2 pieces generated by the two-stage baseline model,
    \item 2 (relatively unknown) human-composed songs: \textit{You Belong to my Heart} (Bing Crosby) and \textit{One Small Photograph} (Kevin Shegog).
\end{itemize}
As stated in Section \ref{sec:leadsheets}, a lead sheet only encodes the basic template of a song, and it ideally needs to be played by a real musician.
We therefore gave all lead sheets to a semi-professional pianist; the pianist stayed true to the sheet music, but was free to create an accompaniment that suited the piece.
In our regards, this evaluation method reflects best how a lead sheet, produced by an AI model, would in practice be used and experienced by musicians and listeners.

The audio clips were presented to each user in randomized order.
For each clip we asked the user to rate on a scale of 1 to 5 how much he likes the piece, if the melody is musically coherent, and whether the piece is composed by a computer (1) or a human (5).
We also asked the user to indicate if he recognizes the piece.
Since each user has his own rating bias and spread \cite{Koren:2009jg,Jin:2004ew}, we converted the ratings for each user to a standardized $Z$ score between $-0.5$ and $0.5$:
\begin{align}
    Z_{c,u} = \frac{R_{c,u} - \mu_u}{\max_{c'} R_{c',u} - \min_{c'} R_{c',u}}.
\end{align}
In this formula, $R_{c,u}$ is the rating of user $u$ for clip $c$, $\mu_u$ is the average rating of user $u$, and $Z_{c,u}$ is the associated standardized score.
Table \ref{table:results} reports the average $\bar{Z}$ score across the audio clips for each of the three questions in the survey, along with the standard deviation.
A negative score means that the ratings are below average overall, and a positive score indicates an overall above-average rating.

We observe that the scores are, by far, better for the two-stage models compared to the unconditioned one-stage model.
This shows that first sampling a harmonic and rhythmic sequence, and conditioning the melody on top of this sequence, is more beneficial than sampling all quantities simultaneously.
Next to this, we also notice that adding the BiLSTM layers improves the score for all three questions.
And although by a small margin, we can conclude that the musical quality improves when the melody generator can look ahead in the harmonic sequence.
When we condition the melody generator on an existing chord and rhythm scheme, it is remarkable that the human-composed and AI-composed songs perform almost on par.
The AI-composed songs are even considered most pleasing.
Related to this observation, 4 participants indicated having recognized a piece from the two-stage model, 5 recognized a piece that was generated based on existing chords, and 3 participants recognized a human-composed song.

Finally, we also want to point out that the standard deviations are very substantial, which shows that there is a high level of disagreement between the reviewers.
It is however interesting to point out that the standard deviation is slightly higher for better performing models.
This might indicate that there is more consensus on what it means for music to sound `badly', but that the definition of `good' music is more subjective and person-dependent.

\begin{table}[t]
\small
\centering
\tabcolsep=0.2cm
\begin{tabular}{l | c c c }
\toprule
\textbf{Model} & \textbf{Pleasing $\bar{Z}$} & \textbf{Coherence $\bar{Z}$} & \textbf{Turing $\bar{Z}$} \\
\hline
One-stage                & $-0.23\pm0.34$ & $-0.24\pm0.31$ & $-0.22\pm0.29$\\
Two-stage, without BiLSTM  & $-0.04\pm0.35$ & $-0.07\pm0.33$ & $-0.09\pm0.33$\\
Two-stage, with BiLSTM     & $-0.01\pm0.36$ & $0.01\pm0.34$  & $-0.02\pm0.34$\\
\hline
Two-stage, with existing chords & $\textbf{0.15}\pm0.37$ & $0.09\pm0.36$    & $\textbf{0.13}\pm0.36$\\
\hline
Human-composed songs        & $0.03\pm0.34$ & $\textbf{0.13}\pm0.36$    & $\textbf{0.13}\pm0.34$\\
\bottomrule
\end{tabular}
\vspace*{1mm}
\caption{Results of the subjective listening experiments. We report averaged $\bar{Z}$-scores for each of the questions, along with the standard deviations.}
\label{table:results}
\end{table}

\section{Conclusion}
\vspace*{-1mm}
We have proposed a two-stage LSTM-based model to generate lead sheets from scratch.
In the first stage, a sequence of chords and rhythm patterns is generated, and in the second stage the sequence of melody notes is generated conditioned on the output of the first stage.
We conducted a subjective listening test of which the results showed that our approach outperformed the baselines.
We can therefore conclude that conditioning helps the quality of the generated music, and that this approach can be explored further in the future.

%
%
%
\vspace*{-2mm}
\bibliographystyle{splncs04}
\bibliography{lib}

\newpage
\appendix
\section{Mode Mapping for Chords}
\label{appendix:modemapping}
In Table \ref{table:modemapping} we show how different chord modes are mapped to one of the following four options: major, minor, diminished or augmented.

\begin{table}[h]
\scriptsize
\centering
\begin{tabular}{l | l || l | l}
\toprule
\textbf{Original mode} & \textbf{Mapped mode} & \textbf{Original mode} & \textbf{Mapped mode} \\
\hline
6	& major	& major-6-9	& major \\
7	& major	& major-7	& major \\
9	& major	& major-9	& major \\
augmented	& augmented	& major-minor	& major \\
augmented-7	& augmented	& minor	& minor \\
augmented-9	& augmented	& minor-11	& minor \\
diminished	& diminished	& minor-13	& minor \\
diminished-7	& diminished	& minor-6	& minor \\
dominant	& major	& minor-7	& minor \\
dominant-11	& major	& minor-7-b5	& diminished \\
dominant-13	& major	& minor-9	& minor \\
dominant-7	& major	& minor-major	& minor \\
dominant-9	& major	& minor-major-7	& minor \\
half-diminished	& diminished	& power	& major \\
major	& major	& sus2	& major \\
major-13	& major	& sus4	& major \\
major-6	& major	& sus4-7	& major \\
\bottomrule
\end{tabular}
\caption{Chord modes are mapped to one of four options.}
\label{table:modemapping}
\end{table}

\section{Rhythm types}
\label{appendix:rhythmtypes}
Table \ref{table:rhythmtypes} provides an overview of the twelve rhythmic figures that are used.

\begin{table}[h!]
\scriptsize
\def\arraystretch{1.4}
\centering
\begin{tabular}{c | c}
\toprule
\textbf{Textual description} & \textbf{Musical symbol} \\
\hline
32nd note & \Zwdr \\
32nd dotted note & \Zwdr\Pu \\
16th note & \Sech \\
8th triplet note & $\overset{\;\;\;\;3}{\AchtBR}$ \\
8th note & \Acht \\
quarter triplet note & $\overset{\quad\;\;3}{\Vier}$ \\
8th dotted note & \Acht\Pu \\
quarter note & \Vier \\
quarter dotted note & \Vier\Pu \\
half note & \Halb \\
half dotted note & \Halb\Pu \\
whole note & \Ganz \\
\bottomrule
\end{tabular}
\caption{The rhythm types that are considered in this paper.}
\label{table:rhythmtypes}
\end{table}

\end{document}